\documentclass[12pt]{article}
\usepackage{amssymb,slashed}
\usepackage{amsmath}
\usepackage[all]{xy}
\newcommand{\ft}[2]{{\textstyle\frac{#1}{#2}}}

\def\rmi{{\mathrm i}}

\newsavebox{\uuunit}
\sbox{\uuunit}
    {\setlength{\unitlength}{0.825em}
     \begin{picture}(0.6,0.7)
        \thinlines
        \put(0,0){\line(1,0){0.5}}
        \put(0.15,0){\line(0,1){0.7}}
        \put(0.35,0){\line(0,1){0.8}}
       \multiput(0.3,0.8)(-0.04,-0.02){12}{\rule{0.5pt}{0.5pt}}
     \end {picture}}

\csname @addtoreset\endcsname{equation}{section}

\def\cU{{\cal U}}

\newtheorem{note}{Remark}[section]
\newtheorem{lemma}{Lemma}[section]
\newtheorem{corollary}{Corollary}[section]
\newtheorem{proposition}{Proposition}[section]

\begin{document}
\begin{titlepage}
\begin{flushright}
UB-ECM-PF 07/34\\
arXiv:0712.0008
\end{flushright}
\vspace{.5cm}
\begin{center}
\baselineskip=16pt {\LARGE Neither black-holes nor regular solitons:\\ a no-go theorem
}\\
\vfill
{\large Alessio Celi
} \\
\vfill
{\small Departament ECM and ICCUB, \\
Institut de Ciencies del Cosmos de la Universitat de Barcelona,\\
            Marti i Franques, 1, E-08028 Barcelona\\
 }
\end{center}
\vfill
\begin{center}
{\bf Abstract}
\end{center}
{\small By  studying the BPS equations for electrostatic and spherically symmetric configurations in $N=2$,
$d=5$ gauged supergravity with vector multiplets and hypermultiplets coupled, we demonstrate that no regular
supersymmetric black-hole solutions of this kind exist. Furthermore, we demonstrate that it is not possible to
construct supersymmetric regular solitons that have the above symmetries. As a consequence the scalar flow
associated to the BPS solutions is always {\it unbounded}. }\vspace{2mm} \vfill \hrule width 3.cm \vspace{3mm}
 {\footnotesize \noindent
e-mail: alessio@ecm.ub.es}
\end{titlepage}
\addtocounter{page}{1}
 \tableofcontents{}

\section{Introduction and result}

Since the advent of the duality between String theory (or, for energies much smaller than the string scale,
Supergravity) on anti-de Sitter (AdS) spaces and supersymmetric conformal field theory residing on the conformal
boundary of AdS \cite{Maldacena:1997re,Gubser:1998bc,Witten:1998qj,Aharony:1999ti}, there has been a renewed
interest in the study of gauged supergravity in various dimensions. A preeminent role has been played in this
contest by special types of solutions like domain walls, black strings and black holes, that admit an
interesting holographic interpretations.


Naturally, a considerable attention has been taken by the five-dimensional supergravity, especially in its
minimal supersymmetric formulation. This theory has eight supercharges ($N = 2$ in the four dimensional
 language) and shares a lot of properties with the theories in four and six dimensions with same amount of
supersymmetry.

The five-dimensional theory is interesting because provides a natural effective set-up to study the extension of
the correspondence beyond the conformal limit and in presence of the less supersymmetry.
 It also favors the construction of phenomenological model with large extra dimensions like
 the Randall-Sundrum scenarios \cite{Randall:1999vf,Randall:1999ee} and it is crucial in any attempt to accommodate them
 with string theory \cite{Ceresole:2001wi}.

 Although a lot of works have been devoted to construct special supersymmetric solutions or/and to classify the possible
 ones, most of them remain unknown, especially for the full-fledge $N=2$, $d=5$ gauged supergravity coupled to matter.
Indeed, even though the program of classifying the theory has been carried out almost\footnote{$N=2$, $d=5$
gauged supergravity with tensor multiplets coupled has not been considered yet.} to completion in various steps
\cite{Gauntlett:2002nw,Gauntlett:2003fk,Gutowski:2004yv,Gutowski:2005id,Celi:2003qk,Bellorin:2006yr,Bellorin:2007yp},
for most of the solutions we have only an implicit construction and the full implication of the classification,
at least when hypermultiplets are coupled, remains to be reveal.

Also for the few classes of BPS solutions that have been studied starting by clever chosen ansatz, it is
difficult to recognize all the consequences that the reach geometric structure of the theory has. A nice example
is given by domain walls. Surprisingly, it was shown only quite recently \cite{Celi:2004st}, under the
inspiration of \cite{Freedman:2003ax}, that supersymmetric domain walls can be ``curved"
\cite{LopesCardoso:2001rt,LopesCardoso:2002ec,Behrndt:2002ee,LopesCardoso:2002ff,Chamseddine:2001hx} only when
hypermultiplets are not trivially coupled.

Nevertheless, the domain walls, together with some of their deformations \cite{Celi:2006pd}, are the only
configurations studied in full generality for all the possible multiplets coupling.

This is not true for black holes, which have been considered in full extent only in presence of vector
multiplets
\cite{Behrndt:1998ns,Behrndt:1998jd,Gutowski:2004ez,Gutowski:2004yv,Chong:2005hr,Kunduri:2006ek,Figueras:2006xx,Behrndt:2004pn,Gauntlett:2004cm}.
More in general, not a lot is known about charged solutions, especially in presence of charged matter
(hypermultiplet) \cite{Cacciatori:2002qx,Cacciatori:2004qm,Liu:2007rv}.

In this work we try to partially cover this gap. We consider all 1/2 BPS electrostatic and spherically symmetric
configurations of the most general $N=2$, $d=5$ supergravity gauged by an abelian gauge group. In our
parametrization, they are described by the metric
\begin{equation}
\label{metricintro} ds^2 =-e^{2v} dt^2 +e^{2w}  dr^2 + r^2 (d\theta^2 +\sin^2 \theta d\phi^2 +\cos^2 \theta
d\xi^2 ),
\end{equation}
where $r$ represents ``physical" radius. The choice of the ansatz is completed by taking the gauge field as the
electrostatic potential. The scalars and all the other fields depends only on $r$, in agreement with the $SO(4)$
symmetry. We analyze the BPS equations for these configurations previously derived in
\cite{Cacciatori:2002qx,Cacciatori:2004qm}.

Some of the solutions are already well known. For example, this class contains the superstar solutions
\cite{Behrndt:1998ns}, singular black hole configurations that can be deformed to non BPS regular black holes
simply by the introduction of a mass term \cite{Behrndt:1998jd}. These configuration are obtained in a
particular vector multiplet model, called ``STU" that is a consistent truncation of $N=8$, $d=5$ gauged
supergravity. As a consequence, the superstars correspond to the consistent reduction of type IIB solutions on
$S^5$ and they can be uplifted back \cite{Cvetic:1999xp} to ten-dimensions. The superstars have received a lot
of attention for their interpretation in terms of Giant graviton excitations \cite{Myers:2001aq}. Finally, in
\cite{Lin:2004nb}, it has been shown that the superstars arise as singular limit of regular 1/2 BPS ``AdS
bubbles'' (see also \cite{Chong:2004ce}). Recently, these singular black hole configurations have been
investigated in the Fake supergravity formulism \cite{Elvang:2007ba}. Furthermore, the analogous of the
superstar solutions in the minimal gauged five-dimensional supergravity can also be interpreted as a solution of
IIB, reduced this time on $Y^{(p,q)}$ \cite{Buchel:2006gb}. As from the ten-dimensional point of view this
solution preserves four supercharges, its non BPS deformations is dual to the quark-gluon plasma of the $N=1$,
$d=4$ quiver gauge theory.

A different example of solutions (which is not asymptotically $AdS_5$) in the family of (\ref{metricintro}),
whose origin in string/M-theory is known, is given by the ``axionic-shift" considered in \cite{Gutperle:2001vw}
and further discussed in \cite{Cacciatori:2002qx}.

Here, we are interested in the understanding of the general properties of the solutions (\ref{metricintro}). In
practise, our aim is draw the identikit of the possible configurations and determine their qualitative behavior
by the analysis of the BPS equations. Due to the symmetry of the problem this can be done without relaying on
the specific features of a particular model but only on the geometric properties of the scalar manifold. By
using such properties we are able to demonstrate that:
\begin{itemize}
\item there are no supersymmetric (extreme) Reissner-Nordstr\"om  in $N=2$ $d=5$ gauged supergravity with
abelian gauging; \item no other regular solution than the supersymmetric vacuum exists, no matter on which
matter multiplets are coupled.
\end{itemize}
These result are obtained in various steps and are based on the analysis of the BPS equations previously derived
in \cite{Cacciatori:2004qm}. We note that when only vector multiplets are present the existence of regular
spherically symmetric BPS black holes has been already ruled out in\footnote{The near-horizon geometry analysis
of \cite{Kunduri:2006uh,Kunduri:2007qy} has been recently compared with the global properties imposed by the
attractor mechanism in \cite{Silva:2007tw}.} \cite{Kunduri:2006uh,Kunduri:2007qy}.

By first, we observe that a necessary condition for a solution to be regular is that the corresponding
superpotential $W$ is a regular and limited function of the scalar manifold. As the BPS equations impose that
$W$ is monotonic decreasing function of $r$, $W'<0$, this is equivalent\footnote{The superpotential without lose
of generality may be taken positive. If, in addition to the regularity in the bulk, we require a regular
behavior at the radial infinity, the superpotential must be also bounded from below, $W_0\ge W\ge W_\infty$,
where $W_\infty\equiv W(r=\infty)>0$. }  to ask that $W_0\equiv W(r=0)$ is finite. As explained  in section
\ref{general} the finiteness of $W$ is not only {\it necessary} but also {\it sufficient}.

Armed of this property, we are able to derive our no-go theorems via {\it reductio ad absurdum}.
Indeed, as a first consequence of the assumption of $W$ finite, we show that regular solutions can only end at the $r=0$.

Then, we prove that such solutions cannot be black holes because $g_{tt}$  and $g_{rr}$  are
constant at the origin.

By considering the BPS equations for the scalars, we argue that the regularity imposes  $W'=0$
at $r=0$. This automatically implies that the vector multiplet scalars should reach a attractor point of $W$, which means
 $\partial_x W_{|_{r=0}} = 0$. Such an attractor point should be infrared attractive. This gives immediately a
 contradiction, due to the famous
No-go theorem \cite{Kallosh:2000tj,Behrndt:2000tr} for very special real geometry, hence regular solutions with
only vector multiplets non trivially coupled
 are  excluded.

By the analysis of the hypermultiplet sector, we first show that, in order to be regular and to satisfy e.o.m.,
a supersymmetric configuration must approach a fixed point of $W$, i.e. $\partial_X W_{|_{r=0}}= 0$. This is
irrespective of the presence of vector multiplets. Then, by considering hypermultiplets only, we demonstrate
that no flow can end at a fixed point for $r=0$: we conclude that the only BPS solution (fully) regular in the
origin is the maximally supersymmetric AdS vacuum. Regarding the generic case, we show that, in opposition to
what happens for the BPS domain wall flows \cite{Ceresole:2001wi} (which present a lot of similarity with the
ones under study), the mixing terms in the superpotential are not sufficient to give an infrared
fixed-point\footnote{This terminology is not fully appropriate in presence of hypermultiplets.}, i.e. attractive
for $r\rightarrow 0$. This is essentially due to the fact that we are studying charged configurations: because
of the coupling with the electric field, the hyperscalars (the scalars of the hypermultiplets) get attracted to
their fixed-point values quicker than the vectorscalars (the scalars of the vector multiplets) in a way that the
behavior of the latter is not affected. We can thus conclude that the superpotential cannot be bounded and no
regular BPS solution different from the supersymmetric $AdS$ vacuum exists.\footnote{It is intriguing to
conjecture that it might exist some link between this result and the absence of Kaluza-Klein $AdS_5$ solitons
proved in \cite{Onemli:2003gg}.} As a consequence, all the solutions that are asymptotically $AdS$ at large $r$
display a superstar-like behavior, irrespectively of the model considered.

\section{A preliminary analysis of the BPS equation}

The class of configurations that we are going to study are 1/2 BPS solutions of the $N=2$, $d=5$ gauged
supergravity, charged under an abelian gauge group. These are bosonic solutions of the e.o.m.\footnote{This has
been explicitly proved in \cite{Cacciatori:2004qm} for configurations under study. A general argument can be
found in \cite{Bellorin:2007yp}.} obtained by solving the first order BPS equations (\ref{sflux}-\ref{5.5_v}).
These follow from the preservation of half of the supersymmetry. In this section we present such equations and
we define all the quantities that are involved in the derivation of the no-go theorems of section \ref{nogo}.
How the BPS equations have been obtained in \cite{Cacciatori:2004qm}, as well as the most important features of
the supergravity theory, is reviewed in the appendix \ref{review}.

The field content of most general $N=2$, $d=5$ gauged supergravity with an abelian gauged group is given by:
\begin{itemize}
\item the supergravity multiplet,
containing the {\em graviton} $e^a_\mu$, two {\em gravitini} $\psi^{\alpha i}_\mu$ and the {\em graviphoton}
$A^0_\mu$; \item $n_H$ hypermultiplets, 
containing the {\em hyperini} $\zeta^A $ with $A=1,2,\dots,2n_H$, and the real scalars $q^X$ with $
X=1,2,\dots,4n_H$ which define a Quaternionic-K\"ahler manifold with  metric $g_{XY}$; \item $n_V$ vector
multiplets,
containing the {\em gaugini} $\lambda^{ia}$, $a=1, \ldots ,n_V$ of spin $\frac 12$, the real scalars $\phi^x $,
 $x=1, \ldots ,n_V$, which define a very special manifold with the metric $g_{xy}$ and $n_V$ {\em gauge vectors} $A^{\hat{ I}}_\mu $,
 $\hat{I} =1\ldots , n_V $. Usually the graviphoton is included by taking $I=0\ldots n_V$. At the same time, the
 very special manifold can be parametrized in terms of $n_V +1$ coordinates $h^I(\phi)$, constrained by $C_{IJK}
 h^I h^J h^K =1$.
\end{itemize}
The corresponding Lagrangian for the bosonic field is \cite{Ceresole:2000jd,Bergshoeff:2004kh}
\begin{eqnarray} {\cal {L}}_{BOS} &=& \frac 12 e\{ R-\frac 12 a_{IJ} F^I_{\mu
\nu} F^{J \mu \nu} -g_{XY} D_\mu q^X D^\mu q^Y -g_{x  y} \partial_\mu \phi^x \partial^\mu \phi^y -2g^2 {\cal V}
(q,\phi) \} \cr & & +\frac 1{6\sqrt 6} \epsilon^{\mu\nu\rho\sigma\tau} C_{IJK} F^I_{\mu\nu}
F^J_{\rho\sigma}A^K_{\tau}. \label{lagr_v}
\end{eqnarray}
The gauge covariant derivative of the hypermultiplet scalars that are charged under the gauged group is
\begin{gather*}
D_\mu q^X=\partial_{\mu} q^X + g A^I_\mu K_I^X(q),
\end{gather*}
where $K_I^X(q)$ are Killing vectors on the quaternionic manifold. ${\cal V} (q,\phi)$ is the scalar potential,
as given in Appendix \ref{review}. A crucial property of the quaternionic geometry is that any isometry $K_I^X$
is associated to a $SU(2)$ triplets of Killing prepotentials $P_{I}^{r}(q)$ $(r=1,2,3)$ via
\begin{equation}
D_{X}P^{r}_{I}= R_{XY}^{r}K^{Y}_{I}, \qquad \Leftrightarrow \qquad \left\{ \begin{array}{c}
  K_{I}^Y=-\frac{4}{3}R^{rYX}D_{X}P^{r}_{I} \\ [2mm]
  D_XP_I^r=-\varepsilon ^{rst}R^s_{XY}D^YP_I^t,
\end{array}\right.
\label{KillingP}
\end{equation}
where $D_{X}$ denotes the $SU(2)$ covariant derivative, which contains an $SU(2)$ connection $\omega_{X}^{r}$
with curvature $ R^r_{XY}$:
\begin{equation}
  D_X P^r= \partial _XP^r+2\,\varepsilon ^{rst} \omega _X^sP^t,\qquad
R^r_{XY}=2\,\partial _{[X}\omega _{Y]}^r+2\,\varepsilon ^{rst}\omega _X^s\omega _Y^t.
 \label{defSU2cR}
\end{equation}
The existence of the prepotential, which in the mathematical literature is known as moment map (see
\cite{Alekseevsky:2001if} and references therein), is essential to guarantee the supersymmetry invariance of the
gauged theory. Furthermore it determines the BPS scalar equations, as we will see.

We can define the {\it dressed} prepotential $P^r$ and the related Killing vector $K^X$ as the linear
combination $P^r\equiv P_I^r h^I$ and $K^X\equiv K^X_I h^I$. Finally, by decomposing $P^r$ in its modulus and
phase we have
\begin{equation}
P^r\equiv \sqrt{\frac32} W Q^r,\ \ \ \ \ \ \ Q^rQ^r=1.
\end{equation}
The scalar function $W$ is known as {\it superpotential}. Let us remark that the prepotential $P^r(q,\phi)$, and
as a consequence the superpotential $W(q,\phi)$, is always well-defined and smooth within the domain of the
scalar manifold \cite{Alekseevsky:2001if} (see also \cite{Behrndt:2001km}).

Now, we are ready to define the ansatz for the solutions of interest. Indeed, the most general electrostatic and
spherically symmetric configuration (up to gauge choice) can be expressed in terms of the metric
\begin{equation}
ds^2 =-e^{2v} dt^2 +e^{2w}  dr^2 + r^2 (d\theta^2 +\sin^2 \theta d\phi^2 +\cos^2 \theta d\xi^2 ),\label{metric}
\end{equation}
and of the electrostatic potentials\footnote{With respect to \cite{Cacciatori:2004qm}, here we are interested in
the generic case $\mu=l^x=0$. The presence of a non trivial $l^x$ is irrelevant for our discussion, as the
scalar fields and the metric are independent of $l^x$.}
\begin{equation}
A_t^I= \sqrt{\frac 32} \Lambda e^v h^I(\phi).
\end{equation}
In accordance with the symmetry, the functions $v$, $w$, $\Lambda$ and the scalars can depend only on $r$. The
requirement that half of the supersymmetry is preserved translates into the projector condition
\begin{equation}
\left(-i \Lambda \gamma_0 \delta_i^j \pm \sqrt{1-\Lambda^2} Q^r {(\sigma_r)_i}^j \gamma_1\right) \epsilon_j
=\epsilon_i.\label{spipro}
\end{equation}
A direct consequence of (\ref{spipro}) is that $\Lambda^2 \le 1$. It is a little more involved to show that
\begin{equation}
\partial_x Q^r = 0.
\end{equation}
As observed in \cite{Cacciatori:2004qm}, the same condition appears in study of BPS flat domain walls
\cite{Ceresole:2001wi}. It has various consequence. From the point of view of the gauging, i.e. of the choice of
the Killing vector $K^X$, it can be seen as a constraint as $\partial_x Q^r = 0$ is not satisfied {\it
everywhere} for a generic linear combination of quaternionic isometries.

On other hands the above condition implies that the BPS solutions depend explicitly on the superpotential $W$
{\it only} and not on the full prepotential $P^r$. A first hint of this fact can be deduces by the for the form
of scalar potential, as it reduces to
\begin{equation}
{\cal V} = -6  W^2  + \frac92  g^{\Lambda \Sigma} \partial_\Lambda W \partial_\Sigma W,
\end{equation}
where the indices run over all the scalars, $\Lambda,\Sigma =1,\dots,n_V+ 4n_H$. The same it happens for the
scalar flow. The complete set of independent BPS equations derived in  \cite{Cacciatori:2004qm} then is:
\begin{gather}
\partial_x Q^r = 0,\label{sflux}\\
1={\Lambda}^2 e^{-2w} \left[ 1 +\frac {r}2
\left( v^\prime + \frac {{\Lambda}^\prime} {\Lambda} \right) \right]^2, \label{5.1_v} \\
e^v =\frac {r}{r_0 \sqrt {1-{\Lambda}^2}}, \label{def_v_v}\\
\sqrt {1-{\Lambda}^2} =\mp g e^w rW, \label{def_w_v}\\
{q^\prime}^Z =\pm 3ge^w \sqrt {1-{\Lambda}^2} \partial^Z W, \label{5.4_v}\\
\phi^{\prime x} =\pm 3ge^w \frac 1{\sqrt {1-{\Lambda}^2}} \partial^x W. \label{5.5_v}
\end{gather}
This tells us that $W$ is the only quantity to be investigated in order to determine the behavior of the BPS
solutions. We will see that qualitatively such behavior is not going to depend too much on the details of the
model. 

We conclude this  review of the BPS equations by noticing that the differential equation for
 the electrostatic potential (\ref{5.1_v}) can be expressed in more clear way. By eliminating $e^w$ due to
 (\ref{def_w_v})
  and defining
\begin{equation}
z=2\frac{1-\Lambda^2}{\Lambda^2} \Rightarrow \Lambda^2 = \frac 1{\frac z2 +1},
\end{equation}
(\ref{5.1_v}) becomes
 \begin{equation}
 \dot z = 6 z - 2 z^2 f^2(y)\label{dz},
 \end{equation}
where $\dot z= \partial_y z$ with $y=\ln r$ and $f \equiv \frac 1 {g r W}$. Treating  $W$ as a function of $y$,
the equation (\ref{dz}) can be formally integrated as
\begin{equation}
z(y) = \frac{e^{6 y}}{2 \int dy e^{6y} f^2}\label{zy},
\end{equation}
or equivalently
\begin{equation}
z(r) = \frac{r^6}{2 \int dr \frac{r^3}{(g W)^2} }\label{zr}.
\end{equation}
The above equation implies for the metric
\begin{eqnarray}
& &e^v= \frac r{r_0} \sqrt{\frac {z+2} z},\\ \label{evz}
& &e^w = \sqrt{\frac z{z+2} f^2}.\label{ewz}
\end{eqnarray}
The ``formal''  integration of $W$ can be made explicit in presence of vector multiplets only (or equivalently
with the hypermultiplet scalars stabilized at constant vevs) because their BPS equation (\ref{5.5_v})  can be
made independent of $z$ ($\Lambda$):
\begin{equation}
\dot \phi^x = - 3 g^{xy} \partial_x \ln W.\label{dphi}
\end{equation}
 On the contrary for the hyperscalars we have
\begin{equation}
\dot q^X = - 3 \frac z{z+2} g^{XY} \partial_Y \ln W.\label{dq}
\end{equation}
The different form of the above equations explains why is relatively easy to find  superstar solutions
analytically (see \cite{Behrndt:1998ns}) while it can be done only for constant $\Lambda$ in presence of
hypermultiplets \cite{Cacciatori:2002qx}. The above equations will be deeply analyzed in the following sections.
As a last comment, we note that the BPS solutions do not depend on the sign of $W$, which is related to the
$\pm$ signs in the equations (\ref{sflux}-\ref{5.5_v}). Without lose of generality $W$ will be taken positive in
the following.

\section{Two no-go theorems}\label{nogo}

\subsection{No BPS black holes}

By definition, $z$ must be positive. This implies that the integration constant in the denominator of
\begin{equation}
z(r) = \frac{r^6}{2 \int_0^r dr' \frac{r^3}{\left(gW(r')\right)^2}+ c}\label{zc},
\end{equation}
 must be positive or the solution will be well-defined  only for open interval $r>r*$. If we are interested in the
  existence of black-hole solutions we need that $e^v \rightarrow 0$ as it approaches the horizon. From the equation
  (\ref{def_v_v}) it is clear that this can occur only at the origin $r=0$, then must be $c\ge 0 $. Another crucial
  requirement for the existence of a regular horizon and then of a black hole is that the scalars have to approach a non
   singular fixed point (i.e. within the domain of the scalar manifold).

This is equivalent to ask that the superpotential admits a well-defined and {\it finite} value on the horizon,
$\lim_{r->0} W = W_0 $. Under this condition we are able to determine the behavior of $z$, and consequently of
the metric for $r\approx 0$. Indeed, its behavior depends dramatically on the value of the integration constant
$c$. For $c\neq 0$, $z\approx r^6$ then $e^v$ is diverging in $r=0$ as $1/r^2$: the solution has a naked
singularity. For $c=0$ instead, the asymptotic   behavior of $z$ is
\begin{equation}
z \approx 2 (gW_0 r)^2, \label{zr1}
\end{equation}
and
\begin{equation}
e^v\approx \frac 1{g W_0 r_0},\ \
e^w= \frac 1{g W_0 r_0} e^{-v} \approx 1.
\end{equation}
This  is sufficient to conclude that NO SUPERSYMMETRY BLACK HOLE EXISTS with an abelian gauging (the non abelian
case is not discussed here and the question remains open). Let us now discuss the existence of a regular
solitons.

We  note that $c=0$ is a necessary condition in order to find regular solutions. Indeed, studying the case of
$c$ negative, which has been left apart, it can be shown that the solution encounters a curvature singularity
for $r\rightarrow r*$. This can be easily understood considering the solution for constant $W=W_0$, that
corresponds to take an asymptotic limit of a generic regular solution. From (\ref{zc}) it follows:
\begin{equation}
z= 2 (g W_0 r)^2 \frac 1{1-(\frac {r*}r)^4}, \ \ \  \ \ \ \ \ \ r*^4= -2 c > 0,
\end{equation}
 and, using  (\ref{evz}-\ref{ewz}),
\begin{eqnarray}
& & e^v= \frac 1{g W_0 r_0} \sqrt{(g W_0 r)^2 + 1 -(\frac{r*}r)^4}, \\
& & e^w= \frac 1{\sqrt{(g W_0 r)^2 + 1 -(\frac{r*}r)^4}},\\
& & A_t= \sqrt{\frac 3z} \frac r{r_0} = \frac 1{g W_0 r_0} \sqrt{\frac 32 (1- (\frac{r*}r)^4)}.
\end{eqnarray}
 Although the metric and the electric potential are regular their derivatives are not: for example the electric field
diverges as $(r-r*)^{-2}$.

On the contrary, taking the limit $r*\rightarrow 0$ we recover the $c=0$ case and the above solution approach
asymptotically the maximally supersymmetric AdS vacuum in spherical coordinates:
 \begin{eqnarray}
& & z= 2 (g W_0 r)^2, \\
& & e^v= \frac 1{g W_0 r_0} \sqrt{(g W_0 r)^2 + 1}, \\
& & e^w= \frac 1{\sqrt{(g W_0 r)^2 + 1}},\\
& & A_t= \sqrt{\frac 32} \frac 1{g W_0 r_0}  .
\end{eqnarray}
Hence, we conclude that only the solutions with $c=0$ can be regular at\footnote{The integration constant $c$
does not play any role for $r \rightarrow \infty$.}  $r=0$. In next session we will show that the only
supersymmetric solution satisfying the hypothesis of a regular flow, i.e. $\exists \lim_{r\rightarrow 0} W =W_0
$ {\it finite}, is AdS vacuum itself.

\subsection{Or constant or unbounded}\label{general}

In what follows we demonstrate that  the assumption of a regular and {\it bounded} flow for the superpotential
$W$ automatically gives that $W$ must be constant. Hence, in any model of the $N=2$, $d=5$ gauged supergravity
with abelian gauging the superpotential associated to a supersymmetric soliton {\it diverges} in the infrared
interior, pointing out that the gauged supergravity, viewed as an effective theory, breaks down in this limit.
In order to explain more clearly the various step of  our proof, we first proceed by analyzing what happens when
hypermultiplets and vector multiplets are considered separately, and afterwards we discuss the generic case.

Before doing so, let us focus on the scalar equation (\ref{dphi}) and (\ref{dq}). As argued in the previous
section, in order to avoid  a ``conical singularity" in the scalar profile their derivative must be zero at the
origin. This automatically implies, from eq.(\ref{dphi}), that at $r=0$ the flow has to reach a stationary point
of $W$ with respect the vector multiplet scalars, $\partial_x W = 0$. At this stage the analogous condition for
the hypermultiplet scalars $\partial_X W = 0$  is not necessary because of the dependence on $z$: the factor
$\frac z{z+2}|_{r\approx 0} \approx \frac z2 \approx (W_0 r)^2$.

\subsubsection*{Just vectors}

When only the vector multiplets are non trivially coupled the problem simplifies greatly and the observation
above together with the properties of special geometry (the main formulas are given in the appendix
\ref{review}) is sufficient to rule out the existence of regular solution. In this case the superpotential $W$
can be written as\footnote{For an abelian gauging, as follows from the consistency condition (\ref{constraint}),
the prepotential $P^r_I$ must not only constant but also parallel. As a consequence the condition for
supersymmetry $\partial_x Q^r = 0$ is automatically satisfied.} $W= W_I h^I(\phi)$, for some constant vector
$W_I$. This implies that the fixed-point condition is $\partial_x W=-\sqrt{\frac 23} W_I h_x^I=0$, or
equivalently $h^I(\phi^*)= W^I/W$. Moreover, for the Hessian of $W$ at $\phi=\phi^*$ it gives
 \begin{eqnarray}
&& \partial^y \partial_x W|_{\phi=\phi^*} = W_I \partial^y (-\sqrt{\frac 23} h_x^I)|_{\phi=\phi^*}=-\sqrt{\frac
23} W h_I(\phi^*) (\partial^y h_x^I)|_{\phi=\phi^*} = \frac 23 W {\delta_x}^y.\cr &&\label{Hv}
\end{eqnarray}
This is essentially  the no-go theorem \cite{Kallosh:2000tj,Behrndt:2000tr} derived studying the properties of
domain wall solutions coupled to vector multiplets (for a compact review of supersymmetric domain walls in $N=2$
supergravity in five dimensions see \cite{Celi:2004st}). Indeed, the above relations imply that there is a
unique fixed-point at $\phi=\phi^*$ and, by expanding the scalar equation (\ref{dphi}) around it, one gets
\begin{equation}
r\partial_r \phi^x \approx - 2 (\phi-\phi^*)^x.
\end{equation}
This tells us that the flow can reach $\phi^*$ only when $r$ is going to infinity. The fixed-point is so called
ultraviolet attractive 
due to the holographic interpretation of scalar equation as the
$\beta$ function of the dual theory. Indeed here $r$ plays the role of the energy scale (as the scale factor
does in the domain wall configurations).

We can thus conclude that the flow is not bounded, $W$ diverges at $r$ and the supersymmetric solutions are not
regular. The solutions that are asymptotic to $AdS_5$ for large $r$ are all superstar-like or $AdS_5$ itself.

\subsubsection*{Just hypers}

As a preliminary step of our analysis, we show that the regularity of the solution forces the condition
$\partial_X W= 0$ when $q'^X=0=W'$, hence for $r=0$. This can be done in two ways. For example it can be shown
by direct calculation, by expanding $W$ around $r=0$ and imposing that the e.o.m are satisfied at the
origin.\footnote{The equations of motion are a consequence of the BPS equations when the scalars are not
constant. Only if there is enhancement of supersymmetry this relation can be extended to hold also at the fixed
point.} Indeed from (\ref{dq}) and (\ref{zr1}) it follows
\begin{equation}
W|_{r\approx 0} = W_0 - \frac 32  W_0 (\partial^X W\partial_X W) r^2 + O[r]^3.
\end{equation}
This expression can be used to compute $z$, $e^v$ and $e^w$ to the next order
\begin{eqnarray}
& & z= 2  W_0 ^2 r^2 (1 -  4 (\partial^X W \partial_X W ) r^2) + O[r]^5, \\
& & e^{2v} = \frac{1}{r_0^2 W_0^2}( 1 + ( W_0^2 + 2 (\partial^X W \partial_X W )) r^2)+O[r]^3, \\
& & e^{2w} =1 - 3 ( W_0 - \partial^X W\partial_X W ) r^2 + O[r]^3.
\end{eqnarray}
As a consequence, it can be shown the Einstein equations become equivalent to $\partial^X W\partial_X W = 0$. We
note that the above result holds also in presence of vector multiplets, because they do not contribute to $W''$
and to the Einstein equations at the leading order at $r=0$.

One can reach the same conclusion by arguing that, as observed in previous section, the regular solution must be
asymptotic to $AdS_5$ for $r\rightarrow 0$. As the flow is supersymmetric, it must be $\partial_X W = 0$ because
the AdS vacuum cannot preserve only a fraction of supersymmetry, hence it must be maximally supersymmetric.
Practically the same argument applies for $r$ going to infinity: the scalars (in this regime hyperscalars and
vectorscalars behave in the same way) can be consistently attracted to a fixed-point if and only if the
derivative of the superpotential with respect to the scalars is zero.

What we have just shown allows us to formulate our question in a way that is easier to answer. By introducing
the parameter $\lambda$ defined as $d\lambda = \frac z{r(z+2)} dr $, we are left to question the regularity of
the flow
\begin{equation} \frac{d q^X}{d\lambda} = -3 \partial^X \log W.\label{dqh}
\end{equation}
 Again, the problem seems equivalent to the analysis of renormalization group flow associated to the flat domain wall
 solutions studied in \cite[2.44]{Ceresole:2001wi}.
 In our contest, a supersymmetry solutions will be regular {\it if and only } the quaternionic scalars are attracted by
 a fixed-point for $r\rightarrow 0$. It is easy to conclude that this cannot be the case as $\lambda$ is finite at $r=0$.
  Indeed by expanding linearly around a fixed point $q^*$ and integrating by part one gets that
  $\log|q-q^*|\propto \lambda$. This leads to a contradiction because the l.h.s. diverges at the fixed-point.

\subsubsection*{Hypers + vectors: the generic case}

Now we are ready to discuss the generic case. The final result will be that the mixing terms arising due to
the presence of both non trivial matter couplings do not help  to escape the no-go theorem derived in the previous
sections. The reason why this happens resides again in the factor $\xi(r)\equiv\frac z{z+2}$ in the hyperscalar equation
  (\ref{dq}). Its presence makes the analysis of the flow, and in particular of the properties of the fixed-points,
  strongly dependent on the two ``regimes": $r\approx\infty$ and   $r\approx 0$. In the former case the analysis
  performed in \cite{Ceresole:2001wi} for the flat domain solutions applies. Indeed, as follows from
\begin{equation}
z(r) = \frac{r^6}{2 \int_0^r dr' \frac{r^3}{\left(gW(r')\right)^2}}\label{z},
\end{equation}
for a limited\footnote{Moreover this implies that $2 (W_\infty r)^2<z< 2 (W_0 r)^2$. The function $z$, which is
related to the value of the electrostatic potential, interpolates monotonically between the two neutral AdS
configurations.}superpotential, $W_0\ge W\ge W_\infty>0$, $z\approx 2 (W_\infty r)^2$. As a consequence,
$\xi(r)\rightarrow 1$ and the scalar equations (\ref{dphi}) and (\ref{dq}) take the form:
\begin{eqnarray}
\dot\varphi^\Lambda &=& - 3 \partial^\Lambda \log W \equiv \beta^\Lambda, \cr
                    &\approx & \partial_\Sigma \beta^\Lambda (\varphi^\Sigma - \varphi_*^\Sigma),
\end{eqnarray}
where we adopt the unified notation for the scalars $\varphi^\Lambda$, $\Lambda =1,\cdots, n_V + 4 n_H$.
 $\varphi_*^\Sigma$ represent the fixed-point values of the scalars for $r=\infty$.
This tells us that any time the matrix
\begin{equation}
\cU_\Sigma {}^\Lambda  \equiv - \, \left. \frac{\partial \beta^\Lambda
}{\partial \varphi^\Sigma }\right|_{\varphi^*} =\left. \frac{3}{W}\, g^{\Lambda
\Xi}\frac{\partial^2 W}{\partial \varphi^\Sigma \partial \varphi ^\Xi
}\right|_{\varphi^*}
 \label{defcU}
\end{equation}
has positive eigenvalues, at least in the direction of the flow, the fixed-point is attractive for $r\rightarrow
\infty$ and the flow can reach it. Thus, in the regime $r\approx \infty$ our problem coincides precisely with
the domain wall one. This is not surprising because the configuration under study (\ref{metric}) can be seen as
a spherical domain wall (the presence of an electric field is a necessary add-on for stabilizing it and solve
the BPS equation): so for large $r$ the two configuration become undistinguishable. Hence, all the BPS solutions
obtained from compact gauging satisfying the properties derived in \cite{Ceresole:2001wi,Alekseevsky:2001if} are
asymptotically $AdS_5$ at the radial infinity.

When $r\approx 0$ the situation is totally different. As discussed in the previous section, now $\xi(r)\approx
(W_0 r)^2$ (again, we assume that $W_0$ is finite). Then, in this region the different nature of a
hypermultiplet scalar with respect to the vector multiplet one clearly emerge. Indeed, the former couples to the
electric field and so it has a very distinct behavior in a charged configuration. From a practical point of
view, as  $\xi(r)\rightarrow 0$, the property of the fixed-point at $r=0$ are not directly related to the
eigenvalue of the Hessian (\ref{defcU}). However, it is still possible to study the stability  linearizing the
flow equations  (\ref{dphi}) and (\ref{dq}).  In order to explain clearly the procedure and  to make evident the
result, we sketch a generic model by considering two scalars $\phi$ and $q$ representing the two different
matter multiplets. After the linearization around a fixed-point, which without lose of generality we may fix at
$\phi^*=q^*=0$, the scalar equations reduced to
\begin{eqnarray}
& & \dot q = \xi(r)(m \phi + M q + O[\varphi]^2), \label{dql}\\
& & \dot \phi = -2 \phi +  m q + O[\varphi]^2,\label{dphil}
\end{eqnarray}
where we used the fact that $U_x^y= 2 {\delta_x}^y$ (cfr. (\ref{Hv})). Here $m$ and $M$ play the role of the
Hessian blocks $-U_y^X$ ($-U^y_X$) and  $-U_Y^X$ respectively. In order to integrate the above system we start
by studying the equation (\ref{dql}).

We remind that $y=\ln r$ and $\dot f \equiv \frac {d f}{d y} = r \frac {d f}{d r}$. Regarding $\phi$ as a
generic function of $y$, we can write:
\begin{equation}
q =
m \exp\left[\frac M2 \int dy \xi \right] \left( \int dy \xi \phi  \exp\left[ - \frac M2 \int dy \xi \right] + c \right).
\end{equation}
 As we are interested in a solution that possibly reaches the fixed-point
 $q^*=0$ for $y\rightarrow  - \infty$ ($r\rightarrow 0$), we choose the integration constant accordingly:
\begin{equation}
q(y) =
m \exp\left[\frac M2 \int_{-\infty}^y dy' \xi(y') \right] \left( \int_{-\infty}^y dy' \xi(y') \phi(y')
 \exp\left[ - \frac M2 \int_{-\infty}^{y'} dy'' \xi(y'') \right] \right).
\end{equation}
Here we need only the leading term in $q$ that corresponds to the first contribution in $\xi$,  as $\xi$ goes to
0 at the fixed point. Thus, we get
\begin{equation} q(y) =  m \int_{-\infty}^y dy' \xi(y') \phi(y').
\end{equation}
We can substitute the above equation in the differential equation for $\phi$:
\begin{equation}
\dot \phi(y) + 2  \phi(y) - m^2 \int_{-\infty}^y dy'\xi(y') \phi(y') = 0. \label{Dphi}
\end{equation}
 This equation can be solved analytically in terms of Bessel functions. Indeed, by defining
 $\chi= \int_{-\infty}^y dy'\xi(y') \phi(y')$ and using that $\xi\propto e^{2y}$ around $r=0$, (\ref{Dphi}) is equivalent
 to
 \begin{equation}
\ddot \chi = m^2 W_0^2 e^{2 y} \chi,
\end{equation}
that is solved by
\begin{equation}
 \chi = c_1 \text{BesselI}[0, |m| W_0 e^y] + c_2 \text{BesselK}[0, |m| W_0 e^y].
\end{equation}
It is easy to check that or $\chi$ is identically zero otherwise its limit  $\lim_{y \to -\infty} \chi$ is always
different from zero for any value of the constants $c_1$, $c_2$ as  $\lim_{y\to-\infty}$BesselI$[0, |m| W_0 e^y] = 1$
and $\lim_{y\to-\infty}$BesselK$[0, |m| W_0 e^y] = \infty$. It follows that $\phi(y)$ diverges at the fixed-point that
is always repulsive.
The calculation proves what the intuition naively suggests: the presence of $\xi$ that is going to zero at the
fixed-point ``cancels out" of the Hessian (\ref{defcU}) all the entrances related to the hypermultiplet sector
and only the repulsive contribution coming from vector multiplets remains. Said in another way, from the vector
multiplet point of view, due to the factor $\xi$, the hyperscalars can be treat as constant: thus we reduce
effectively to the case when only vector multiplets are coupled.

It remains to check that the result we got from the toy-model (\ref{dql}-\ref{dphil}) remains unchanged when we consider
 the full-fledge scalar equations. Indeed, this is case. By first we observe that only the scalars whose derivatives
  receive contribution at the linear level from the mixing terms are affected by the presence of both the multiplets
  at the same time.  Hence, only the flows in these directions have to be discussed as they are the only left over by the
  no-go theorem of the previous sections. This means that we can always choose a base in which we consider only the
  scalars $q^{\tilde X}$, $\phi^{\tilde x}$, with $
  {\tilde X},{\tilde x}=1,\dots, l\equiv$ rank$[U_X^x]$, and
  $U_{\tilde X}^ {\tilde x} = m_{(\tilde x)} \delta_{\tilde X}^{\tilde x}$. Furthermore, as also the matrix
  $U_{\tilde X}^ {\tilde Y}$ is symmetric, we can diagonalize it by a rotation on the
  $q^{\tilde X}$, $U_{\tilde X}^ {\tilde Y}= M_{(\tilde Y)} \delta_{\tilde X}^ {\tilde Y}$, and at
  the same time keep $U_{\tilde X}^ {\tilde x}$ in the diagonal form by antirotating the $\phi^{\tilde x}$. Hence, we
  end up with $l$ copies of the system (\ref{dql}-\ref{dphil}),
\begin{eqnarray}
& & \dot q^i = \xi(r)(m_{(i)} \phi^i + M_{(i)} q^i + O[\varphi]^2), \\
& & \dot \phi^i = -2 \phi^i +  m_{(i)} q^i + O[\varphi]^2,
\end{eqnarray}
where $m$ and $M$ are replaced by the eigenvalues $m_{(i)}$ and $M_{(i)}$. As a final remark, let us stress that
our conclusions are not affected by non linear terms. These enter the game when the linear contribution around
the fixed-point is zero. Indeed, this is obviously never the case when only vector multiplets are coupled. On
the contrary, this is possible in the hypermultiplets sector. However, when the vector multiplets are not
coupled the analysis of (\ref{dqh}) remains unchanged as $\lambda $ in any case should be divergent while is
not. Regarding the generic case, as the $q$'s can scale {\it atmost}\footnote{This fact can be proven by using
the l'H\^opital theorem in order to calculate the limit of the ratio $\frac q\phi$, $\lim \frac q\phi = \lim
\frac {\dot q}{\dot\phi}$. This is essentially a consequence of $\xi\to0$.} as the $\phi$'s, we can trust the
Taylor expansion of the scalar derivatives, in particular of the $\dot\phi$'s. Thus, the only relevant mixing
term that can appear in the $\dot\phi$'s (and as a consequence in the $\dot q$'s) is the linear one discussed
above. If this is absent, we can directly conclude that the $\phi$'s have a repulsive behavior as
$\dot\phi\approx -2 \phi$.

Therefore, we conclude that the hypothesis of a superior bound $W_0$ for the superpotential is contradicted by
the absence of an attractive fixed-point for $r\to 0$. This prove that the existence of regular electrostatic
and spherically symmetric BPS soliton is ruled out  for {\it any} model of $N=2$, $d=5$ gauged supergravity with
matter coupling.

\section{Conclusion}

Any supersymmetric solution different from the vacuum ends with a singularity in the interior of the spacetime.
In practice all the BPS solutions that preserve strictly half of the supersymmetry behave qualitatively like a
superstar \cite{Behrndt:1998ns}  around $r\approx 0$, while at the radial infinity they may be asymptotical to
(maximal supersymmetric) $AdS_5$  or they may display a runaway behavior as the analytical solution presented in
\cite{Cacciatori:2002qx}. This common behavior at the origin can be interpreted as a sort of attractor
mechanism. Indeed only the ultraviolet (UV) seems sensible to the details of the supergravity model.

The finding of this universal singular behavior at $r=0$ has various consequences and suggests new questions.
First of all, for an asymptotically $AdS_5$ solution it maps via holography in a universal behavior of RG flow
of the dual theory, supposed it exists. Indeed, by adapting the argument of \cite{Girardello:1998pd} (see also
\cite{Freedman:1999gp}), we can construct a c-function $C(r)$ that at large $r$ (UV) will correspond to the
central charge of the conformal theory  on $\mathrm{R}\times S^3$ \cite{Henningson:1998gx}. Now, if we trust the
c-theorem even if the solutions is not reaching any horizon in the infrared, we conclude that the RG flow always
ends with zero central charge, as $C(r)$ is proportional (as in the domain wall case) to $1/W^3$.

A related issue is given by the interpretation of the asymptotically $AdS_5$ BPS solutions in terms of type IIB
string theory. In the work of \cite{Lin:2004nb} it was shown that the superstars correspond to a singular limit
of regular AdS bubbles. In spite of the fact that generically IIB solutions can not be consistently truncated to
the $N=2$, $d=5$ solutions, it is very intriguing to conjecture that behind this universal singular behavior a
new relation between pentadimensional and ten-dimensional configurations may be hidden. If this is the case, the
corresponding ten-dimensional regular solutions should preserve at least 1/8 of the 32 supercharges. As the BPS
equations for these class of configurations have been recently obtained by \cite{Gava:2006pu,Gava:2007kr}(see
also \cite{Ahn:2005vk}), it should be possible to check whether the interpretation of the superstars as 1/2 BPS
AdS bubbles can be generalized and extended, at least in some special cases.

It would be very interesting to see whether our no-theorem extend to fake supergravity theories. Indeed, our
result does not depend on the precise details of the model but on the relation between the superpotential $W$
and the electric field $F_{rt}$, exemplified in the function $z$, as this relation determines the properties of
the scalar flow. These last properties should only depend on the supersymmetric invariance at the linear order
\cite{Elvang:2007ba}. The only point where the full-fledge supersymmetric invariance seems to play a crucial
role is in the determination of the special geometry and, as a consequence, of the Hessian of the
superpotential. For this reason, we may naively suppose that also fake BPS electrostatic and spherically
symmetric black holes are ruled out while solitons could exist. In order to discuss this issue, the fake
supergravity formalism\footnote{A different Fake supergravity formalism has been developed in
\cite{Ceresole:2007wx,Andrianopoli:2007gt,Lopes Cardoso:2007ky} with the aim of studying the attractor mechanism
of extreme non BPS black hole in $N=2$ supergravity in four and five dimensions.} of \cite{Elvang:2007ba} must
be extended to include charged matter. An indication of how this can be done should come from the relation
between $N=2$, $d=5$ supergravity and fake theory discussed in \cite{Celi:2004st}. This issue is currently under
investigation.

Another question raised by our work, is what is going to change if we consider configurations with the same
symmetries but in presence of non-abelian gauging. In particular would be very interesting to consider the
existence of black-holes. Due to the technical difficulties that such a study implies, the use of the
classification achieved in \cite{Bellorin:2007yp} seems crucial.

\medskip
\section*{Acknowledgments.}
\noindent The author thanks Jorge G. Russo for the useful discussions and for proofreading part of this
manuscript. He is graceful to Roberto Emparan, Diederik Roest, and Marcello Ortaggio for discussions.

AC is supported by the ``Ministerio de Educaci\'on y Ciencia", Spain through the ``Programa Juan de la Cierva".
This work is supported in part by the European Community's Human Potential Programme under contract
MRTN-CT-2004-005104 `Constituents, fundamental forces and symmetries of the universe', by the Spanish grant MCYT
FPA 2004-04582-C02-01 and by the Catalan grant CIRIT GC 2005SGR-00564.

\appendix

\section{Spherically symmetric configurations $N=2$ $d=5$ gauged supergravity}\label{review}

\subsection{Five-dimensional $N=2$ gauged supergravity}
We start by  recalling some of the  most important   features of five-dimensional, $\mathcal{N}=2$ gauged
supergravity theories. Further technical details can be found in the original references
\cite{Gunaydin:1984bi,Gunaydin:1985ak,Gunaydin:1999zx,Ceresole:2000jd,Bergshoeff:2004kh}.

The matter multiplets that can be coupled to $5D$, $\mathcal{N}=2$ supergravity are vector, tensor and
hypermultiplets: as we are interested to abelian gauging we end up with vector multiplets and hypermultiplets
only.

The $(n_V)$ scalar fields of $n_{V}$ vector multiplets parameterize a ``very special'' real manifold
$\mathcal{M}_{\rm VS}$, i.e., an $\left(n_{V}\right)$--dimensional hypersurface of an auxiliary
$(n_{V}+1)$-dimensional space spanned by coordinates $h^{I}$  $(I=0,1,\ldots, n_{V}+1)$ :
\begin{equation}
\mathcal{M}_{\rm VS}=\{ h^I\in \mathbb{R}^{(n_V+1)}: C_{IJK}h^I h^J h^K =1\}, \label{defVS}
\end{equation}
where the constants $C_{IJK}$ appear in a Chern-Simons-type coupling
 of the Lagrangian.
On $\mathcal{M}_{\rm VS}$, the embedding coordinates $h^I$  become functions of the physical scalar fields,
$\phi^{x}$ $(x=1,\ldots,n_{V})$. The metric on the very special manifold is determined via the equations
\begin{eqnarray}
  &&g_{xy}=h_x^I\,h_{yI},\qquad h_x^I\equiv
  -\sqrt{\ft32}\,\partial _xh^I, \qquad
  h_I\equiv C_{IJK}h^Jh^K,
\qquad h_{Ix}\equiv \sqrt{\ft32}\,\partial _xh_{I},\nonumber\\
  &&h^Ih_J+h_x^I\,g^{xy}\,h_{yJ}=
  \delta^{I}_{J}, \qquad h^{J}h_I=1,\qquad h^Ih_{Ix}=0.
\label{identVS}
\end{eqnarray}
Some useful relations are
\begin{eqnarray}
 h_{Ix;y} & = & \sqrt{\ft23}\left( h_I g_{xy}+T_{xyz}h^z_I\right)\, ,  \nonumber\\
 h^I{}_{x;y} & = & -\sqrt{\ft23}\left( h^I
 g_{xy}+T_{xyz}h^{Iz}\right)\, , \label{hIxy}\end{eqnarray}
where `;' is a covariant derivative using the Christoffel connection calculated from the metric $g_{xy}$, with
 \begin{equation}
 T_{xyz}  \equiv  C_{IJK}h^I_xh^J_yh_z^K\, .
 \label{Txyz}
\end{equation}
Furthermore, we can define the auxiliary metric $a_{IJ}$ of tangent space $(h^I,h^I_x)$, which determine the
kinetic term of the gauge field in supergravity Lagrangian, as $a_{IJ} =h_Ih_J + h_I^xh_{Jx}$.

The scalars  $q^X$ $(X=1,\ldots 4n_{H})$ of  $n_{H}$ hypermultiplets, on the other hand, take their values in a
quaternionic-K{\"a}hler manifold $\mathcal{M}_{\rm Q}$ \cite{Bagger:1983tt}, i.e., a manifold of real dimension
$4n_H$ with holonomy group contained in $SU(2)\times USp(2n_H)$. We denote the vielbein on this manifold by
$f_X^{iA}$, where $i=1,2$ and $A=1,\ldots,2n_{H}$ refer to an adapted $SU(2)\times USp(2n_{H})$ decomposition of
the tangent space. The hypercomplex structure is $(-2)$ times the curvature of the $SU(2)$ part of the holonomy
group\footnote{In fact, the proportionality factor includes the Planck mass and the metric, which are implicit
here.}, denoted as $R^{rZX}$ $(r=1,2,3)$, so that the quaternionic identity reads
\begin{equation}
  R^r_{XY}R^{sYZ}=-\ft14\,\delta ^{rs}\,\delta _X{}^Z
   -\ft12\,\varepsilon ^{rst}\,R^t_X{}^Z.
 \label{quaterid}
\end{equation}

Besides these scalar fields, the bosonic sector of the matter multiplets also contains
 $n_{V}$  vector fields from the $n_{V}$ vector multiplets.
Including the graviphoton, we thus have a total of $(n_{V}+1)$ vector fields, $A_{\mu}^{I}$
$(I=0,1,\ldots,n_{V})$, which can be used to gauge up to $(n_{V}+1)$ isometries of the quaternionic-K{\"a}hler
manifold $\mathcal{M}_{\rm Q}$ (provided such  isometries exist). As we restrict to abelian gauging only, there
is no action on the ``very special'' real manifold $\mathcal{M}_{\rm VS}$.

The quaternionic Killing vectors $K_{I}^{X}(q)$ that generate the isometries on $\mathcal{M}_{\rm Q}$ can be
expressed in terms of the derivatives of $SU(2)$ triplets of Killing prepotentials $P_{I}^{r}(q)$ $(r=1,2,3)$
via
\begin{equation}
D_{X}P^{r}_{I}= R_{XY}^{r}K^{Y}_{I}, \qquad \Leftrightarrow \qquad \left\{ \begin{array}{c}
  K_{I}^Y=-\frac{4}{3}R^{rYX}D_{X}P^{r}_{I} \\ [2mm]
  D_XP_I^r=-\varepsilon ^{rst}R^s_{XY}D^YP_I^t,
\end{array}\right.
\label{KillingP}
\end{equation}
where $D_{X}$ denotes the $SU(2)$ covariant derivative, which contains an $SU(2)$ connection $\omega_{X}^{r}$
with curvature $ R^r_{XY}$:
\begin{equation}
  D_X P^r= \partial _XP^r+2\,\varepsilon ^{rst} \omega _X^sP^t,\qquad
R^r_{XY}=2\,\partial _{[X}\omega _{Y]}^r+2\,\varepsilon ^{rst}\omega _X^s\omega _Y^t.
 \label{defSU2cR}
\end{equation}

 The prepotentials
satisfy the constraint
\begin{equation}\label{constraint}
\frac{1}{2}R_{XY}^{r}K_{I}^{X}K_{J}^{Y}- \varepsilon^{rst}P_{I}^{s}P_{J}^{t}
+\frac{1}{2}f_{IJ}{}^{K}P_{K}^{r}=0,
\end{equation}
where $f_{IJ}{}^{K}$ are the structure constants of the gauge group.

In the following, we will frequently switch between the above vector notation for $SU(2)$-valued quantities such
as $P_{I}^{r}$, and  the usual
   $(2\times 2)$   matrix notation,
\begin{equation}
P_{Ii}{}^j\equiv \rmi\,\sigma_{ri}{}^jP_{I}^{r}  .
 \label{matrixVector}
\end{equation}

The general Lagrangian of the gauged supergravity  theory under consideration is
\begin{eqnarray} {\cal {L}}_{BOS} &=& \frac 12 e\{ R-\frac 12 a_{IJ} F^I_{\mu
\nu} F^{J \mu \nu} -g_{XY} D_\mu q^X D^\mu q^Y -g_{x  y} \partial_\mu \phi^x \partial^\mu \phi^y -2g^2 {\cal V}
(q,\phi) \} \cr & & +\frac 1{6\sqrt 6} \epsilon^{\mu\nu\rho\sigma\tau} C_{IJK} F^I_{\mu\nu}
F^J_{\rho\sigma}A^K_{\tau}, \label{lagrapp}
\end{eqnarray}
where
\begin{gather*}
D_\mu q^X=\partial_{\mu} q^X + g A^I_\mu K_I^X(q).
\end{gather*}
Then the variations of the fermions for abelian gauge symmetry ${U(1)}^{n_V+1}$ reduce to:\\
 for the gravitini
\begin{eqnarray} \delta_\epsilon \psi_{\mu i} &=& \partial_\mu \epsilon_i +\frac 14 \omega^{ab}_\mu \gamma_{ab} \epsilon_i
-\partial_\mu q^X p_{Xi}^{\quad j} \epsilon_j + g A^I_\mu P_{Ii}^{\ j} \epsilon_j \cr &+&\frac i{4 \sqrt 6}
(\gamma_{\mu \nu \rho} -4g_{\mu \nu} \gamma_\rho )h_I F^{I\nu \rho } \epsilon_i - \frac i{\sqrt 6} g h^I
P_{Ii}^{\ j} \gamma_\mu \epsilon_j =0; \label{BPSgrav_v} \end{eqnarray} for the gaugini \begin{eqnarray}
\delta_\epsilon \lambda_i^{X} = -\frac i2 \gamma^a \partial_a \phi^{X}\epsilon_i
 -\frac i2 \gamma^a g A_a^I K_I^{X} +\frac 14
h_{I}^{X} \gamma^{ab} F^{I}_{ab} \epsilon_i -g h^{X I} P_{Ii}^{\quad j} \epsilon_j =0 \label{gaugini}
\end{eqnarray}
for the hyperini \begin{eqnarray} \delta_\epsilon \zeta^A =\left[ -\frac i2 \gamma^a \partial_a q^X -\frac i2
\gamma^a g A_a^I K_I^X +gh^I \frac {\sqrt 6}4 K_I^X \right] f_{Xi}^A \epsilon^i =0 \label{hyperini}
\end{eqnarray}
We are now ready to discuss the BPS equations for electrostatic and spherically symmetric configuration.

\subsection{BPS equation for spherically symmetric configurations}
Now, we review the main steps of derivation of the BPS equations as originally done
\cite{Cacciatori:2002qx,Cacciatori:2004qm}. First of all we select the form of the ansatz. We are looking for
electrostatic spherically symmetric  solutions that preserve half of the $N=2$ supersymmetries. We choose
metric,  which is $SO(4)$-symmetric, as
\begin{equation}
ds^2 =-e^{2v} dt^2 +e^{2w}  dr^2 + r^2 (d\theta^2 +\sin^2 \theta d\phi^2 +\cos^2 \theta d\xi^2
).\label{metricapp}
\end{equation}
with the function $v$ and $w$, as well as the scalars, that only depend on the spacetime coordinate $r$. The
same can be achieved for the gauge fields by fixing the gauge in order to have only the $A^I_t$ component
different from zero. Thus we can write
\begin{equation}
A^I_t =\sqrt {\frac 32} a^I e^v,
\end{equation}
where $a^I$ are generic functions of $r$. However, as a vector, it can be decomposed on the flow as $a^I = a h^I
+ l^xh_x^I$. It is shown in \cite{Cacciatori:2004qm} that the choice $l^x=0$ is always compatible with the
supersymmetry preservation. Furthermore, the presence of a non trivial $l^x$ cannot affect the result of section
\ref{nogo}, as the scalar fields and the metric are independent of $l^x$. For this reason, we consider only the
case $l^x=0$.

The next step is to substitute the above ansatz in the supersymmetry transformations in order to find a
covariantly constant spinor $\epsilon$. The BPS equations will arise as consistency conditions. A smart
procedure is to consider first the matter transformations. Due to the ansatz, the gaugini (\ref{gaugini}) and
hyperini (\ref{hyperini}) variations become respectively:
\begin{eqnarray} \delta_\epsilon \lambda^x_i &=& \left[ -i \phi^{\prime x} e^{-w} \gamma_1
\delta_i^{\ j} -2 i g h^{Ix} P_I^s (\sigma_s)_i^{\ j} \right. \cr & & \left. + \sqrt {\frac 32} e^{-w} h_I^x
\left(v^\prime a^I + {a^I}^\prime \right) \gamma_{01} \delta_i^{\ j} \right] \epsilon_j =0 \label{gaugini_v}
\end{eqnarray}
and
\begin{eqnarray} \delta_\epsilon \zeta^A &=& f_{iX}^A \left[ - i q^{\prime X} e^{-w} \gamma_1 + i \sqrt {\frac
32} g a^I K_I^X \gamma_0 +\sqrt {\frac 32}  g h^I K_I^X \right] \epsilon^i =0, \label{hyperini_v}
\end{eqnarray}
where we use the notation $\chi'\equiv \partial_r \chi $. The projector content of the above equation can be
written as
\begin{eqnarray}
\left(-i \Lambda \gamma_0 \delta_l^{\ k} +f^r (\sigma_r )_l^{\ k} \gamma_1 \right)\epsilon_k =\epsilon_l,
\end{eqnarray}
where the consistency imposes $\Lambda^2+ f^rf^r =1$.

By the use of the identities of the quaternionic manifold, the hyperini equation turns out to be equivalent (see
also \cite{Celi:2003qk,Celi:2006pd}) to
\begin{eqnarray}
&& f^r= \pm \sqrt{1-\Lambda^2} Q^r \\
&& {q^\prime}^Z =\pm 3ge^w \sqrt {1-{\Lambda}^2} \partial^Z W, \\
&& \sqrt {1-{\Lambda}^2} =\mp g e^w rW.
\end{eqnarray}
As the projector condition coming from the hyperini equation must coincide with the one obtained from the
gaugini, we get the constraint $\partial_xQ^r=0$ and the BPS equation for the $\phi$'s:
\begin{eqnarray}
\phi^{\prime x} =\pm 3ge^w \frac 1{\sqrt {1-{\Lambda}^2}} \partial^x W.
\end{eqnarray}
Furthermore, by consistency the electrostatic potential is fixed to be
\begin{equation}
A_t^I= \sqrt{\frac 32} \Lambda e^v h^I,
\end{equation}
as $a^I= \Lambda h^I$. Finally, the integrability condition of the gravitini turns out to be compatible with the
above equations if
\begin{equation}
e^v =\frac {r}{r_0 \sqrt {1-{\Lambda}^2}}.
\end{equation}

Let us summarizing the results. The preservation of half of the supersymmetry, associated to the projector
condition
\begin{equation}
\left(-i \Lambda \gamma_0 \delta_i^j \pm \sqrt{1-\Lambda^2} Q^r {(\sigma_r)_i}^j \gamma_1\right) \epsilon_j
=\epsilon_i,
\end{equation}
is equivalent for an electrostatic and spherically symmetric solution to the BPS equations
\begin{gather}
\partial_x Q^r = 0,\\
1={\Lambda}^2 e^{-2w} \left[ 1 +\frac {r}2
\left( v^\prime + \frac {{\Lambda}^\prime} {\Lambda} \right) \right]^2,  \\
e^v =\frac {r}{r_0 \sqrt {1-{\Lambda}^2}}, \\
\sqrt {1-{\Lambda}^2} =\mp g e^w rW, \\
{q^\prime}^Z =\pm 3ge^w \sqrt {1-{\Lambda}^2} \partial^Z W, \\
\phi^{\prime x} =\pm 3ge^w \frac 1{\sqrt {1-{\Lambda}^2}} \partial^x W.
\end{gather}
As explicitly shown in \cite{Cacciatori:2004qm}, the above equations are sufficient to ensure the equations of
motion.

\end{document}